\documentclass{article}
\usepackage{arxiv}
\usepackage{amsmath}
\newcommand{\RomanNumeralCaps}[1]
    {\MakeUppercase{\romannumeral #1}}
\newcommand{\mynum}[1]{\textcircled{\raisebox{-0.9pt}{#1}}}
\usepackage{url,hyperref,lineno,microtype,subcaption}
\usepackage[onehalfspacing]{setspace}

\usepackage[utf8]{inputenc} 
\usepackage[T1]{fontenc}    
\usepackage{hyperref}       
\usepackage{url}            
\usepackage{booktabs}       
\usepackage{amsfonts}       
\usepackage{nicefrac}       
\usepackage{microtype}      
\usepackage{lipsum}		
\usepackage{graphicx}
\usepackage{natbib}
\usepackage{doi}

\title{Phase-Aberration Correction in Shear-wave Elastography Imaging Using Local Speed-of-Sound Adaptive Beamforming}

\author{Bhaskara Rao Chintada \\
	Computer-assisted Applications in Medicine\\
	ETH Zurich, Switzerland\\
	\And
    Richard Rau \\
	Computer-assisted Applications in Medicine\\
	ETH Zurich, Switzerland\\
	\And
    Orcun Goksel\thanks{corresponding author} \\
	Computer-assisted Applications in Medicine\\
	ETH Zurich, Switzerland \\
	Department of Information Technology\\
	Uppsala University, Sweden\\
	\texttt{ogoksel@ethz.ch}\\
	}
\renewcommand{\shorttitle}{Phase-Aberration Correction in Shear-wave Elastography Imaging}

\begin{document}
\maketitle

\begin{abstract}
Shear-wave Elastography Imaging (SWEI) is a noninvasive imaging modality that provides tissue elasticity information by measuring the travelling speed of an induced shear-wave. 
It is commercially available on clinical ultrasound scanners and popularly used in the diagnosis and staging of liver disease and breast cancer.
In conventional SWEI methods, a sequence of acoustic radiation force (ARF) pushes are used for inducing a shear-wave, which is tracked using high frame-rate multi-angle plane wave imaging (MA-PWI) to estimate the shear-wave speed (SWS).
Conventionally, these plane waves are beamformed using a constant speed-of-sound (SoS), assuming an a-priori known and homogeneous tissue medium.
However, soft tissues are inhomogeneous, with intrinsic SoS variations.
In this work, study the SoS effects and inhomogeneities on SWS estimation, using simulation and phantoms experiments with porcine muscle as an abbarator, and show how these aberrations can be corrected using local speed-of-sound adaptive beamforming.
For shear-wave tracking, we compare standard beamform with spatially constant SoS values to software beamforming with locally varying SoS maps.
We show that, given SoS aberrations, traditional beamforming using a constant SoS, regardless of the utilized SoS value, introduces a substantial bias in resulting SWS estimations.
Average SWS estimation disparity for the same material was observed over 4.3 times worse when a constant SoS value is used compared to that when a known SoS map is used for beamforming. 
Such biases are shown to be corrected by using a local SoS map in beamforming, indicating the importance of and a need for local SoS reconstruction techniques.

\end{abstract}


\section{Introduction}
Shear-wave elastography Imaging (SWEI) is a noninvasive imaging technique that maps shear-wave speed (SWS) in tissues. 
Conventionally SWEI is performed in two steps:
First in the vicinity of soft tissue to be imaged, a remote 'push' is generated using acoustic radiation force (ARF) to induce shear-waves. Second, these shear-waves are observed using ultrasound imaging to capture lateral shear-wave travel speed~\cite{sarvazyan1998shear}, to relate this to the underlying tissue shear modulus.
Supersonic Shear-wave Imaging (SSI) aims to estimate shear-wave map of the soft tissue with high Signal-to-Noise Ratio (SNR), by generating shear-waves with the constructive interference of multiple pushes along the depth direction, while tracking the shear waves using ultrafast plane wave imaging (PWI)~\cite{bercoff2004supersonic} typically at 10\,000 frame-per-second.   
SWS measurements have been used in many clinical applications including the diagnosis and staging of diseases in the liver, breast, and kidney~\cite{sarvazyan2011elasticity,deffieux2015investigating,tanter2008quantitative}.
SWEI methods usually assume soft tissues are acoustically homogeneous with a nearly-constant SoS, both for the generation of ARF pushes and for the beamforming of PWI in spatially tracking shear-waves.
However, soft tissues are acoustically inhomogeneous, which
may thus introduce artifacts in SWS estimation.
\cite{shi2011phase}~compared SWS and SWS dispersion values measured \textit{in-vivo} and \textit{ex-vivo} on three porcine livers to study the confounding effects of porcine skin/fat/muscle on SWS measurements.
\cite{carrascal2016phase}~studied phase aberration and ultrasound attenuation effects on SWS and shear-wave frequency domain characteristics.
\cite{huang2016phase}~reported that SWS estimation errors due to phase aberrations originate mainly from tracking rather than the ARF push generation.
In this work, we aim to correct such errors in SWS estimation.
Several data acquisition sequences have been proposed in the literature to mitigate phase aberration effects in shear-wave tracking~\cite{montaldo2009coherent,amador2016improvement,espindola2017shear}. 
However, these methods target mediums with slight variations in SoS, whereas in a clinical setting, several layers with largely varying thicknesses and SoS may exist between the ultrasound transducer and the location of measurement.
To alleviate large phase aberration effects, knowing the local SoS distribution would be essential.

Several methods have been introduced in the literature to estimate local SoS distributions in soft tissues, mostly aiming for diagnostic purposes that may be afforded by SoS contrast.
These methods are known as ultrasound computed tomography (USCT) methods.
Conventional USCT systems are based on submerging the target anatomical structure in a water bath, which is equipped with a large number of cylindrically/spherically positioned transducer elements at known locations~\cite{greenleaf1981clinical,gemmeke20073d}.
Such transmission USCT systems have great potential for \textit{in-vivo} breast cancer screening.
As it is naturally beneficial to develop SoS imaging to be compatible with existing conventional ultrasound transducers in order to avail several logistic advantages of commercial transducer arrays, also for SoS imaging in the clinics.
Time-of-flight recordings together with a passive acoustic reflector~\cite{krueger1998limited,sanabria2018speed} or minute misalignments between images viewed from different angles~\cite{jaeger2015computed,sanabria2018spatial,rau2019speed} were used for tomographic reconstruction of SoS.
Given SoS maps, delays to any spatial location can also be calculated to correct for aberrations caused by SoS inhomogeneities, these delays can be used for beamforming, called SoS-adaptive beamforming, which was shown to increase the resolution of B-mode imaging in~\cite{rau2019aberration}.
Herein, we hypothesize that SoS-adaptive beamforming may be used for mitigating affects on SoS estimation, e.g.\ via hindering displacement estimation in shear-wave tracking and/or confounding the apparent speed of shear wavefront in consecutive frames due to aberrations when constant and incorrect SoS values, using simulations and \textit{ex-vivo} phantoms designed to introduce aberration effects. 

\section{Methods}
\subsection{Data acquisition}
\begin{figure}
\begin{center}
\includegraphics[width=1\textwidth]{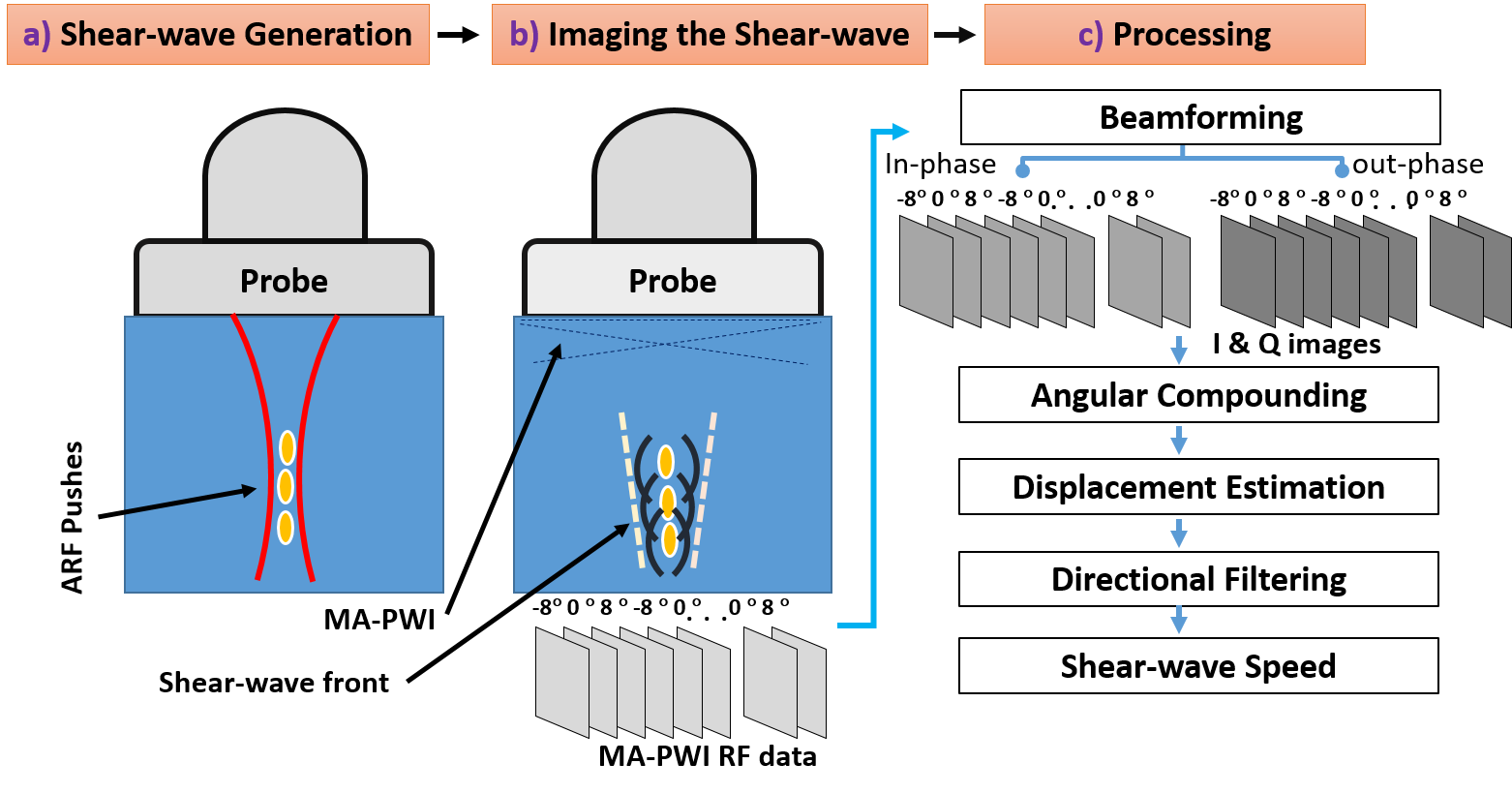}
\end{center}
\caption{Schematic of Shear-wave Elasticity Imaging: (a)~Shear-wave generation; (b)~Tracking the shear-wave using Multi-Angle Plane-Wave Imaging~(MA-PWI); and (c)~Data processing to estimate the shear-wave speed (SWS).}\label{fig:Data_processing}
\end{figure}
In this study we employ shear-waves induced via ARF using the supersonic shear-wave imaging technique~\cite{bercoff2004supersonic}, which generates a quasi-cylindrical shear-wave front.
We use five consecutive high-intensity ARF pushes of each 200\,$\mu$s duration, at five axially separated foci with a separation of 2.5\,mm in depth, as illustrated in Figure\,\ref{fig:Data_processing}(a). 
Laterally propagating shear-wave is then tracked utilizing multi-angle plane-wave imaging (MA-PWI) at a high frame-rate of 10\,K frames-per-second, using three angled plane waves at (-8$^{\circ}$,0$^{\circ}$,8$^{\circ}$) as depicted in Figure\,\ref{fig:Data_processing}(b). 
Both the acquisition sequences for generating ARF and for tracking shear-wave imaging using MA-PWI were programmed in a research ultrasound machine (Verasonics, Seattle, WA, USA) with a 128-element linear-array transducer (Philips, ATL L7-4) operated at 5\,MHz center frequency. 

\subsection{Processing pipeline}
We follow the SWEI data processing pipeline as illustrated in Figure\,\ref{fig:Data_processing}(c).
The raw radio frequency (RF) signals for the MA-PWI frames were collected and recorded during the experimental runtime, in order to software-beamform these using alternative approaches retrospectively.
After beamforming, a moving window of three PWs (with distinct angles) are coherently compounded to increase SNR~\cite{montaldo2009coherent}. 
Note that this does not reduce the frame-rate.
From the compounded frames, axial displacements are estimated using a 2D Loupas autocorrelation method~\cite{loupas1995axial}.
These are then directionally filtered to separate the left and right propagating waves~\cite{manduca2003spatio} along the depth axis $z$. 
Using these shear-wave axial displacement profiles, 
there are several methods in the literature to estimate the SWS~\cite{mclaughlin2006shear,wang2010improving,rouze2010robust,song2014fast}.
In this study we used the 1D SWS estimation method of~\cite{song2014fast}, which we observed to be a robust estimator during preliminary tests for our earlier work \cite{chintada2020nonlinear}.
In this method, SWS at a position $(x,z)$ in axial and lateral axes, respectively, is computed using a model of wave propagation, i.e.\ via multiple normalized cross-correlations (NCC) between $w$ consecutive displacement profile pairs, each $p$-pixels apart, within a lateral window $\left[x+\frac{w}{2}, x-\frac{w}{2}\right]$ at depth $z$.
The final SWS is then the  NCC-weighted average of maximum correlating profile delays, normalized by the wave travel distance equivalent to $p$-pixels.
For a detailed implementation of this method, please refer to~\cite{song2014fast}.

Parameter $w$ provides an averaging effect, increasing SNR but hampering resolution by smoothing out spatial variations as a tradeoff.
Setting parameter $p$ is also a tradeoff; small values allow comparing displacement waveforms of sufficient similarity (i.e.\ reducing dissimilarity due to dispersion, etc.), while large values allow increasing precision thanks to waveforms with larger shifts in time.
Parameters $p$ is set based on the minimum expected SWS in the given medium, which we assumed to be 1.5\,m/s in our study.
In this study we accordingly set $w=4$ and $p=8$. 

\subsection{Beamforming}
For the beamforming of MA-PWI seen above, we employ delay-and-sum using dynamic aperture with an f-number of 0.75 and using the time-delays computed by one of the two options below:
First, as the conventional approach, a constant SoS (e.g., an average value from the literature for the imaged anatomy) is used for computing time delays.
When the target anatomy is unknown, a typical value of choice is 1540\,m/s, which is also the SoS for the CIRS phantom used.
Given the aberrator and different compositions in imaging field, we tested several constant background values in this work.
Second, as an alternative approach, we used known local SoS maps, annotated manually from the images, for aberration correction in beamforming as in~\cite{rau2019aberration}.
Delays $\tau_i$ from all 128 transducer elements to all the locations on an N\textsubscript{x} $\times$ N\textsubscript{z} beamforming grid can be computed given local SoS $\sigma_{map}$ using the relationship
\begin{equation}
  \tau_i = L_i\sigma_{map}, i=[1,2,..., N_x \times N_z \times 128]  
\end{equation}
where $L_i$ are the rows of a path matrix as in~\cite{rau2019aberration}.
Note that such path matrix for beamforming only needs to be computed once given a transducer geometry and beamforming grid (depth), so it can also be precomputed.
Furthermore, given the fixed image composition across all MA-PWI frames, time delays $\tau_i$ are fixed among these frames.

\section{Experiments}

\begin{figure}
\begin{center}
\includegraphics[width=0.5\textwidth]{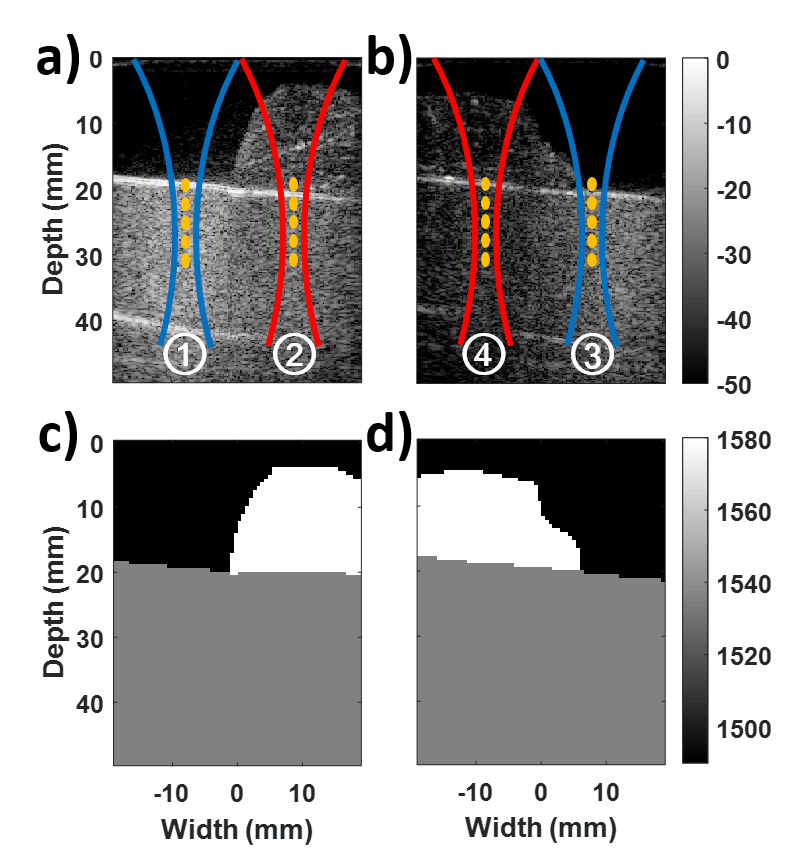}
\end{center}
\caption{Illustration of experimental configurations, (a) B-mode of water and porcine muscle layers placed on CIRS phantom with acoustic radiation force push markings and its SoS map ($\sigma_{map}$) (b) for experimental configurations \mynum{1} and \mynum{2}. Similarly, B-mode and ground truth SoS maps of experimental configurations \mynum{3} and \mynum{4}.}\label{fig:exp_configurations}
\end{figure}

For experiments we used a standard elasticity phantom, CIRS Elasticity QA (Norfolk, VA, USA), with a manufacturer-declared SoS of 1540\,m/s.
To mimic aberrations from real tissues, we placed an \textit{ex-vivo} porcine muscle on the CIRS phantom, and placed the US transducer above it so that the aberration source is in one half of the US imaging field-of-view, as seen in Figure\,\ref{fig:exp_configurations}(a).
We filled water on top of the phantom as acoustic coupling medium.
Before starting experiments, we ensured that the temperature of the water and hence the muscle sample within are stabilized at 22.4$^{\circ}$C, tracked using a thermometer for control.

In total, we used four experimental settings, with two phantom configurations, as shown in Figure\,\ref{fig:exp_configurations}.
Utilizing experiments with the same material but with different combinations allowed us to study beamforming with different SoS effects on SWS estimation, while differentiating effects on the imaging stage from the ARF push path.
We created two phantom configurations by changing the location, volume, and cross-sectional profile of the porcine muscle sample overlaid on the CIRS phantom, as shown in Figure\,\ref{fig:exp_configurations}(a),(b).
The two configurations show some variation in acoustic propagation characteristics and thereby help us identify any systematic effects and demonstrate repeatability of any findings.
For each configuration, we then conducted SWS estimation experiments with ARF pushes ($i$)~using the first 64 elements of the transducer over the water layer side, and ($ii$)~using the last 64 elements over the muscle sample side.
In both cases, all 128 elements were used for imaging, i.e.\ to receive the signals during MA-PWI (although not all elements contribute to all beamformed points due to dynamic focusing).

To be used in aberration correction with local SoS maps, we manually segmented the three material regions from the images of the two phantom settings, as seen in Figure\,\ref{fig:exp_configurations}(c),(d).
We set their SoS as follows:
For the CIRS phantom, we used the declared value.
For the water, we computed the SoS given our monitored temperature using~\cite{bilaniuk1993speed}.
For the muscle sample, we measured the time-of-flight (ToF) from the strong reflection at the phantom-muscle interface (see the white lines within markings \mynum{2} and \mynum{4} in Figure\,\ref{fig:exp_configurations}(a),(b)), and observed its shift with and without the muscle sample being there, to get a relative estimate of the muscle (given its observed thickness) with respect to the known water SoS.
Accordingly, the water, CIRS phantom, and the muscle sample  were set to have SoS values of 1490\,m/s, 1540\,m/s, and 1580\,m/s, respectively.
For each acquisition, to generate a plane-wave in 2D (quasi-conical wave in 3D) a total of 5 consecutive ARF pushes were conducted at depths \{20.0, 22.5, 25.0, 27.5, 30.0\}\,mm.
In our preliminary experiments with homogeneous medium, SWS estimation was observed to be minimally affected by the SoS choice in ARF push Tx delays.
Therefore, we used a fixed SoS of 1540\,m/s for all ARF push Tx delay calculations.
SWS was estimated using~\cite{song2014fast} after beamforming the MA-PWI using constant SoS values of \{1480, 1490, ..., 1590\}\,m/s as well as using the local SoS map~\cite{rau2019aberration}. 
For each of the four experimental configurations and for any given beamforming setting, the acquisition process described above was repeated 5 times and the resulting five SWS maps were point-wise averaged to increase the SNR.
These averaged SWS maps are used below to compare and quantify effects from SoS aberrations. 

\section{Results and Discussion}

Below we first show and analyze the results from the experimental setting \mynum{1} in detail, and then summarize the findings from all four settings comparatively.

\subsection{Detailed analysis of a sample experimental setting}
\begin{figure}
\begin{center}
\includegraphics[width=\textwidth]{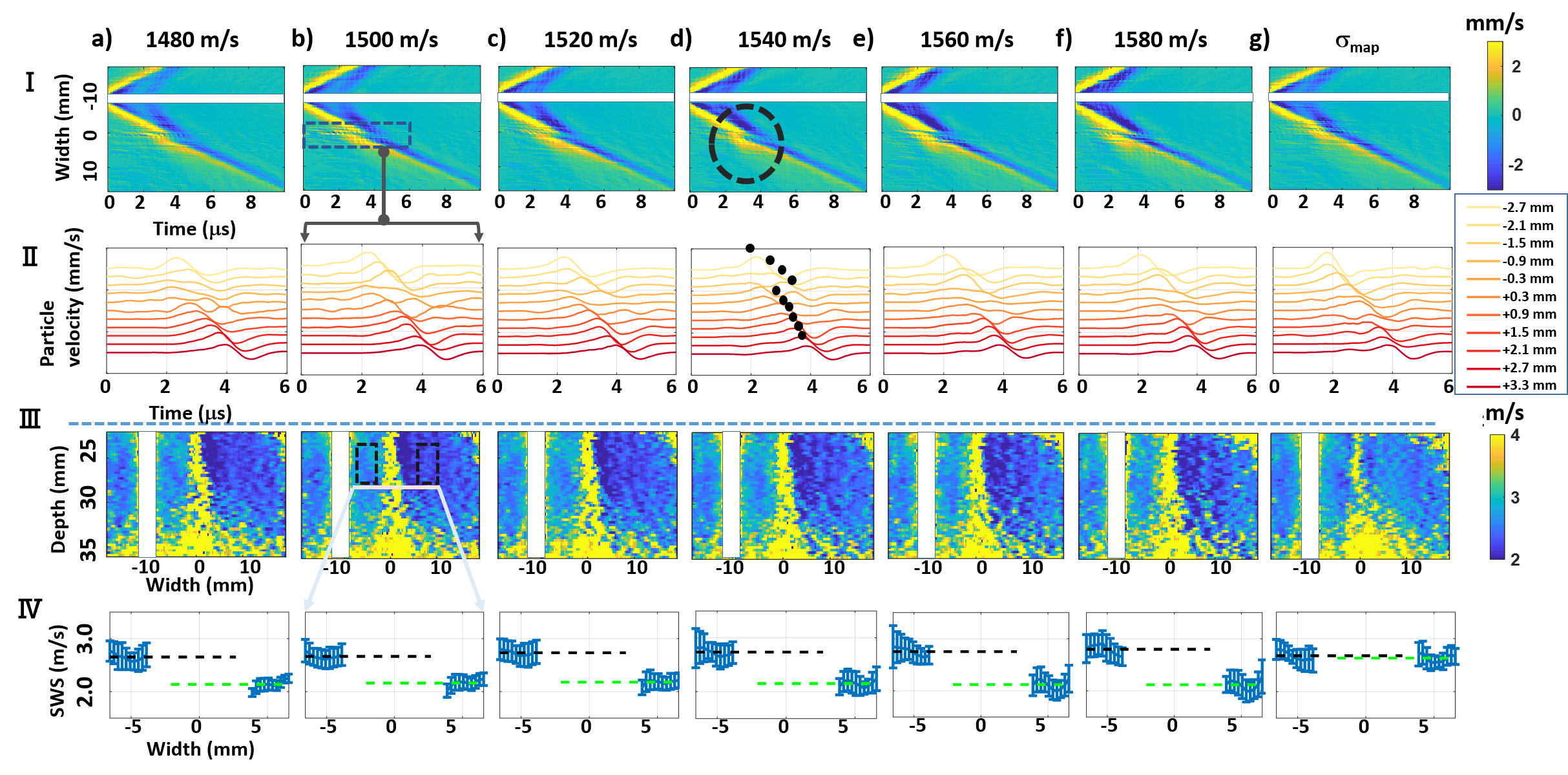}
\end{center}
\caption{Experimental results. \RomanNumeralCaps{1}. Shear-wave axial velocity time and propagation distance profiles; \RomanNumeralCaps{2}. Axial velocity profiles corresponding to rectangular region in Figure\,\ref{fig:results}(\RomanNumeralCaps{1},b)\,; \RomanNumeralCaps{3}. resulting shear-wave speed (SWS) maps; and \RomanNumeralCaps{4}. SWS profiles over selected rectangular regions-of-interest (ROI),  with the errorbars representing the standard deviations of SWS values along the vertical axis, with the dashed lines marking the average of their means, black for the left side and green for the right.
Each column shows results for beamforming the same data acquisition with different speed-of-sound (SoS) maps, (a-f) for constant SoS maps from 1480\,m/s to 1580\,m/s, and (g) using a spatially-varying SoS map $\sigma_{map}$.
We do not report the results, by masking them in white, within vertical bands around the lateral push locations where the shear-wave travels before the tracking starts.} 
\label{fig:results}
\end{figure}

To illustrate the effect of aberration when beamformed with different SoS values, the axial velocity profiles at a selected depth of 25\,mm (as the middle push location, with a relatively ideal shear wavefront), averaged over an axial window of 3\,mm around the selected depth 25\,mm, are shown for the experimental configuration \mynum{1} in Figure\,\ref{fig:results}(\RomanNumeralCaps{1}).
These profiles are shown as a function of time across the phantom width, and they were computed from the estimated displacements, for the beamforming of which we first used different constant SoS values (Figure\,\ref{fig:results}(\RomanNumeralCaps{1},a-f)).
It is seen that during the propagation of the shear wave, it becomes distorted along a vertical band around the mid-line of the phantom ($x=0$), irrespective of the constant beamforming SoS value.
Such distortion band is where part of the received echos used in the beamforming go through the muscle, while the other half through the water, with substantially different SoS values, therefore the beamforming being largely out-of-phase. 
Indeed, the severity of the distortion does not seem to change given the beamforming SoS value and even for the manufacturer-declared 1540\,m/s, such distortion is observed as shown circled in
Figure\,\ref{fig:results}(\RomanNumeralCaps{1},d).
To further illustrate this effect, ribbon plots in Figure\,\ref{fig:results}(\RomanNumeralCaps{2}) zoom in on such distortion regions (delineated with a rectangle in Figure\,\ref{fig:results}(\RomanNumeralCaps{1},b)) to visualize the time-domain characteristics of the axial velocity profiles, i.e.\ the observed shape and time shifts between consecutive axial velocity profiles.
One can see in these profiles that the velocity profiles in the distortion bands lose the coherent wavefront appearance that they otherwise had.

Figure\,\ref{fig:results}(\RomanNumeralCaps{3}) displays the resulting SWS maps obtained within a rectangular field-of-view (FoV) of size 14$\times$35.2\,mm ([22,36]\,mm axially and [-17.6,17.6]\,mm laterally).
Such FoV was selected where we know that the (CIRS) phantom has a constant homogeneous SWS, given the phantom construct and the survey B-mode image in Figure\,\ref{fig:exp_configurations}.
In the estimated SWS maps, severe distortions can be observed around the mid vertical lines ($x=0$) due to the large SoS inhomogeneity, i.e.\ large variations in backscatter wavefront paths considered in beamforming, which introduce aberrations in the shear-wave particle velocity profiles, regardless of the depth where the shear wavefront is observed.
Therefore SWS cannot be reliably estimated along such narrow vertical bands.
Besides this immeasurable region, a striking observation is that the SWS estimated on the left and right sides of such band are not equal (seen as different blue hues), although the estimations can be made with relatively high SNR.
This is also observed in the velocity profiles in Figure\,\ref{fig:results}(\RomanNumeralCaps{2}) where the wavefronts on either side of the distortion band exhibit a mismatch in gradient (which leads the local SWS estimate), as illustrated by marking their automatically-selected peak locations in Figure\,\ref{fig:results}(\RomanNumeralCaps{2},d).
This is not only a random noise or distortion, and indicates a systematic bias in SWS estimates due to an aberrator and the conventional beamforming process, which may affect and alter any diagnostic decision in a clinical setting.
Throughout the rest of this paper, we systematically analyze and further quantify such effects, and study if such bias can be reduced or removed.

To better illustrate the differing SWS values due to aberration, Figure\,\ref{fig:results}(\RomanNumeralCaps{4}) presents some statistics of SWS values within rectangular ROIs of 3.0$\times$4.5\,mm that are placed equidistant from the middle of the imaging width, as marked in Figure\,\ref{fig:results}(\RomanNumeralCaps{3},b).
In these subfigures, the values around the distortion bands were not plotted as these explode the axis range and also have large very variations due to low estimation accuracy.
From the SWS mean values, marked with dashed lines of black and green in Figure\,\ref{fig:results}(\RomanNumeralCaps{4}), one can see that the SWS estimations on either side differ largely, despite both sides having relatively low standard deviations.
Since we know from the phantom construct that the given FoV has a constant homogeneous SWS, one would expect for the SWS values to match.
Note that even if the beamforming SoS errors would change the arrival time of echos to the transducer (and therefore potentially degrade image quality and resolution), one would not expect this to affect the speed estimation of laterally propagating shear-waves, and certainly not in a systematic or an easily explainable way.
Further of note is that this systematic bias occurs regardless of the chosen constant SoS value for beamforming, seen across the columns (a-f).
To the best of our knowledge, we are unaware of a systematic reporting or study of this in the literature.

For beamforming the MA-PWI data using a spatially-varying SoS map in Figure\,\ref{fig:results}(g), we observe in (\RomanNumeralCaps{1},g) and \RomanNumeralCaps{3}g that the distortion regions are substantially minimized.
From the continuous appearance of the velocity profiles (peaks) in (\RomanNumeralCaps{2},g) as well as the constant color hue in the SWS map in (\RomanNumeralCaps{3},g), we can see that using the knowledge of the local SoS variation in beamforming can indeed remove the SWS estimation bias.
This is further illustrated in (\RomanNumeralCaps{4},g) where SWS means within ROIs on both sides have statistically indifferent SWS estimations, showing that beamforming with local SoS maps can remedy such SWS bias and problem.

\subsection{Comparison of beamforming for all experimental settings}

To validate the above observations by controlling for the medium that the ARF push beams have travelled as well as any systematic effect from the direction of the phantom/muscle placement, we analyzed the results from four different experimental configurations:
We tested two phantom configurations, by changing the muscle to the left (in the experiments \mynum{1} and \mynum{2}) and the right side (in \mynum{3} and \mynum{4}).
In each configuration, we applied the ARF push either through the  water layer (in \mynum{1} and \mynum{3}) or through the porcine muscle (in \mynum{2} and \mynum{4}).  
Since the configurations were nearly symmetric, we used the same ROI boxes above, fixed for all the experimental configurations.
We then report the mean and standard deviations of SWS values within each ROI and over 5 repetitions in Table\,\ref{results_tab}.
Note that given the experimental configuration, an ROI is either nearer or farther from the ARF push location, and an ROI is imaged either mostly through water or through muscle.
To help isolate any effects from different factors, we use the following terminology in reporting the results below:
We denote the SWS results in ROI nearer and farther from the ARF push with $S$ and $S'$.
For results in ROI imaged through water (W) and muscle (M) we use the relevant letter in the subscript, i.e.\ $S_\text{W}$ and $S_\text{M}$.
For instance, in Figures\,\ref{fig:results}(\RomanNumeralCaps{4},d), $S_\text{W}$$>$$S'_\text{M}$ although in (\RomanNumeralCaps{4},d) with local SoS map $S_\text{W}$$\approx$$S'_\text{M}$.
We tabulate the SWS estimations for different beamforming SoS values in Table\,\ref{results_tab}
\renewcommand{\tabcolsep}{2.5pt}
\begin{table*}
  \caption{
  The measured SWS values in ROI nearer ($S$) and farther ($S'$) from the ARF when imaged through water (i.e.\ $S_\text{W})$ and muscle (i.e.\ $S_\text{M}$). $D=|S-S'|$ is the mean absolute difference between the two ROIs in each experimental setting and SoS and all units are in [m/s].}
 
  \label{results_tab}
  \begin{center}
    \small
    \begin{tabular}{|l||c|c|c||c|c|c||c|c|c||c|c|c|}
      \textbf{SoS} & \multicolumn{3}{c||}{\textbf{Exp \mynum{1}} - ARF through W} & \multicolumn{3}{c||}{\textbf{Exp \mynum{2}} - ARF through M} & \multicolumn{3}{c||}{\textbf{Exp \mynum{3}} - ARF through W} & \multicolumn{3}{c|}{\textbf{Exp \mynum{4}} - ARF through M} \\ \hline
     & $S_\text{W}$ & $S'_\text{M}$ & $D$ & $S_\text{M}$ & $S'_\text{W}$ & $D$ & $S_\text{W}$ & $S'_\text{M}$ & $D$ & $S_\text{M}$ & $S'_\text{W}$ & $D$ \\ \hline 
     1480& 2.65\footnotesize{$\pm$0.21} & 2.14\footnotesize{$\pm$0.14} & 0.51 & 2.07\footnotesize{$\pm$0.17}& 2.56\footnotesize{$\pm$0.12}  & 0.49 &	2.79\footnotesize{$\pm$0.28} & 2.22\footnotesize{$\pm$0.18}  & \underline{0.57} & 2.32\footnotesize{$\pm$0.32} & 2.64\footnotesize{$\pm$0.22} & \underline{0.32}\\
     1490& 2.66\footnotesize{$\pm$0.23} &2.16\footnotesize{$\pm$0.14}& \underline{0.50} & 2.08\footnotesize{$\pm$0.20}& 2.56\footnotesize{$\pm$0.12}  & 0.48 &	2.85\footnotesize{$\pm$0.32} & 2.27\footnotesize{$\pm$0.17}  & 0.58 & 2.33\footnotesize{$\pm$0.19} & 2.69\footnotesize{$\pm$0.22} & 0.36\\
     1500& 2.67\footnotesize{$\pm$0.18} &2.16\footnotesize{$\pm$0.14}& 0.51 & 2.09\footnotesize{$\pm$0.20}& 2.55\footnotesize{$\pm$0.11}  & 0.46 &	2.86\footnotesize{$\pm$0.24} & 2.27\footnotesize{$\pm$0.15}  & 0.59 & 2.35\footnotesize{$\pm$0.26} & 2.70\footnotesize{$\pm$0.15} & 0.35\\
     1510& 2.70\footnotesize{$\pm$0.20} &2.16\footnotesize{$\pm$0.13}& 0.54 & 2.10\footnotesize{$\pm$0.20}& 2.55\footnotesize{$\pm$0.13}  & 0.45 &	2.93\footnotesize{$\pm$0.28} & 2.26\footnotesize{$\pm$0.14} & 0.67 & 2.34\footnotesize{$\pm$0.20} & 2.75\footnotesize{$\pm$0.17} & 0.41\\
     1520& 2.74\footnotesize{$\pm$0.21} &2.17\footnotesize{$\pm$0.15}& 0.57 & 2.14\footnotesize{$\pm$0.22}& 2.56\footnotesize{$\pm$0.11}  & 0.42 &	2.97\footnotesize{$\pm$0.23} & 2.28\footnotesize{$\pm$0.14} & 0.69 & 2.41\footnotesize{$\pm$0.24} & 2.79\footnotesize{$\pm$0.16} & 0.38\\
     1530& 2.72\footnotesize{$\pm$0.21} &2.15\footnotesize{$\pm$0.14}& 0.57 & 2.14\footnotesize{$\pm$0.21}& 2.56\footnotesize{$\pm$0.12}  & 0.42 &	2.97\footnotesize{$\pm$0.20} & 2.29\footnotesize{$\pm$0.16}  & 0.68 & 2.44\footnotesize{$\pm$0.29} & 2.79\footnotesize{$\pm$0.17} & 0.35\\
     1540& 2.75\footnotesize{$\pm$0.21} &2.16\footnotesize{$\pm$0.19}& 0.59 & 2.11\footnotesize{$\pm$0.21}& 2.53\footnotesize{$\pm$0.14}  & 0.39 &	2.99\footnotesize{$\pm$0.19} & 2.27\footnotesize{$\pm$0.19}  & 0.72 & 2.44\footnotesize{$\pm$0.30} & 2.83\footnotesize{$\pm$0.16} & 0.39\\
     1550& 2.75\footnotesize{$\pm$0.22} &2.14\footnotesize{$\pm$0.21}& 0.61 & 2.16\footnotesize{$\pm$0.24}& 2.55\footnotesize{$\pm$0.12}  & 0.39 &	3.06\footnotesize{$\pm$0.24} & 2.28\footnotesize{$\pm$0.19}  & 0.78 & 2.47\footnotesize{$\pm$0.31} & 2.84\footnotesize{$\pm$0.14} & 0.37\\
     1560& 2.76\footnotesize{$\pm$0.21} &2.12\footnotesize{$\pm$0.22}& 0.64 & 2.17\footnotesize{$\pm$0.24}& 2.55\footnotesize{$\pm$0.15}  & 0.38 &	3.09\footnotesize{$\pm$0.23} & 2.31\footnotesize{$\pm$0.23}  & 0.78 & 2.49\footnotesize{$\pm$0.31} & 2.88\footnotesize{$\pm$0.16} & 0.39\\
     1570& 2.80\footnotesize{$\pm$0.22} &2.12\footnotesize{$\pm$0.21}& 0.68 & 2.18\footnotesize{$\pm$0.27}& 2.56\footnotesize{$\pm$0.17}  & 0.38 &	3.14\footnotesize{$\pm$0.24} & 2.29\footnotesize{$\pm$0.32}  & 0.85 & 2.49\footnotesize{$\pm$0.36} & 2.91\footnotesize{$\pm$0.21} & 0.50\\
     1580& 2.81\footnotesize{$\pm$0.22} &2.11\footnotesize{$\pm$0.21}& 0.70 & 2.22\footnotesize{$\pm$0.29}& 2.56\footnotesize{$\pm$0.21}  & 0.34 &	3.19\footnotesize{$\pm$0.27} & 2.30\footnotesize{$\pm$0.29}  & 0.89 & 2.52\footnotesize{$\pm$0.34} & 2.94\footnotesize{$\pm$0.22} & 0.42\\
     1590& 2.84\footnotesize{$\pm$0.23} &2.15\footnotesize{$\pm$0.25}& 0.69 & 2.28\footnotesize{$\pm$0.36}& 2.55\footnotesize{$\pm$0.22}  & \underline{0.27} &	3.23\footnotesize{$\pm$0.33} & 2.30\footnotesize{$\pm$0.36} & 0.93 & 2.56\footnotesize{$\pm$0.46} & 2.97\footnotesize{$\pm$0.32} & 0.41\\ \hline
     $\sigma_\textrm{map}$& 2.68\footnotesize{$\pm$0.20} &2.63\footnotesize{$\pm$0.18}& \textbf{0.05} & 2.63\footnotesize{$\pm$0.20}& 2.55\footnotesize{$\pm$0.12}  & \textbf{0.08} & 2.30\footnotesize{$\pm$0.17} &	2.42\footnotesize{$\pm$0.38} & \textbf{0.12} & 2.51\footnotesize{$\pm$0.22} & 2.28\footnotesize{$\pm$0.46} & \textbf{0.23}\\ \hline
    \end{tabular}
    \normalsize
  \end{center}
\end{table*}
\nopagebreak
\begin{figure}
    \centering
    \includegraphics[width=.96\columnwidth]{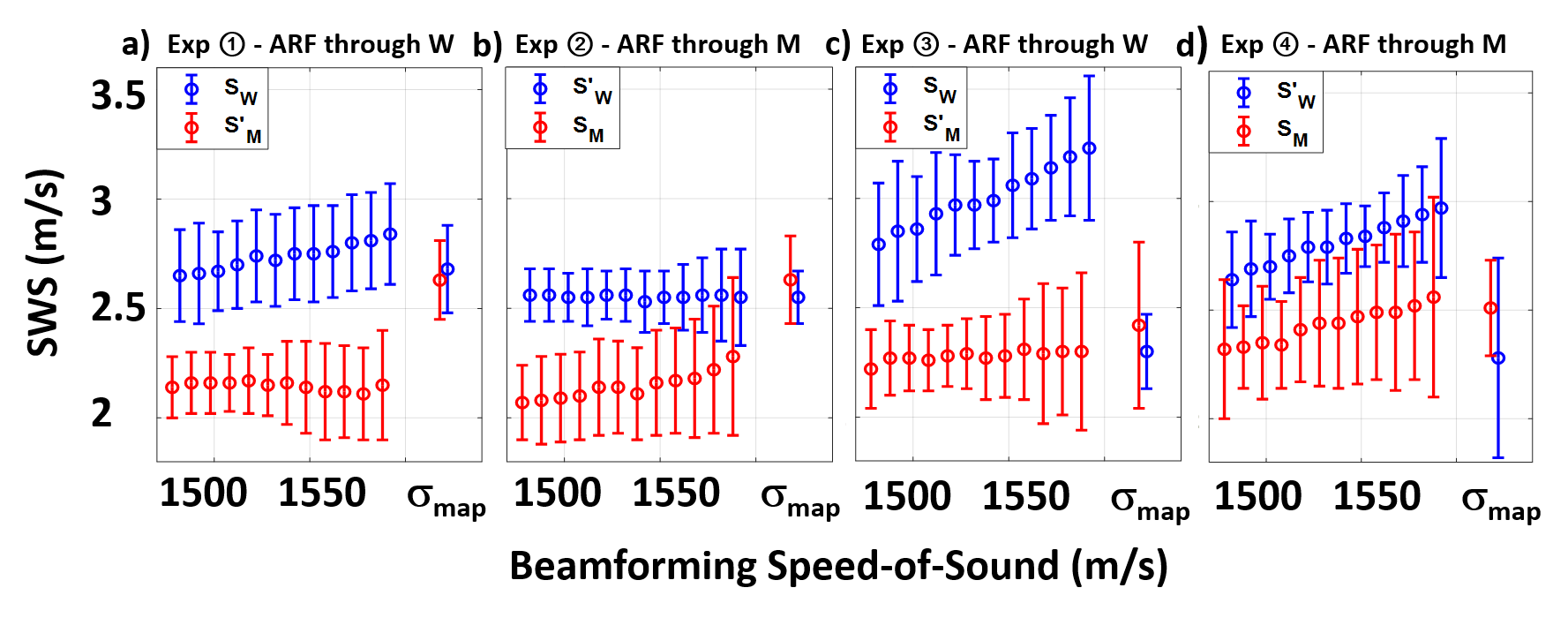}
    \caption{Visual summary of Table\,\ref{results_tab} with SWS values obtained from the two ROIs in each experimental setting when beamformed with constant SoS values and $\sigma_{map}$.}
    \label{fig:table_plot}
\end{figure}
From the table it is seen that the SWS estimations are lower in ROIs on the muscle side of the distortion band, in comparison to the water side.
This is true irrespective of the ARF push going through the muscle or the water, and again irrespective of the muscle side being closer to or farther from the ARF push.
We also report the mean absolute difference (disparity) $D=|S-S'|$ between the two ROIs in each experimental setting and SoS.
It is seen that for constant SoS values for beamforming, SWS differences $D$ are quite large.
For visual comparison of these values, we have plotted SWS values obtained from the mentioned ROIs for each experimental configuration in Fig.\,\ref{fig:table_plot}.
Ideally this SWS disparity should be near zero, since it is the exact same material being measured in both ROIs.
Indeed, the average disparity of all four experiments for a typical (also CIRS-declared) SoS of 1540\,m/s is 0.52\,m/s, which is over 4.3 times higher than the average disparity (0.12\,m/s) when a local SoS map $\sigma_{map}$ is used for beamforming the MA-PWI data.
Furthermore, even for using a best constant SoS value to minimize such difference for each individual experiment (underlined in the table), the average is 0.42\,m/s, which is again 3.5 times higher than that by using the local SoS map.
Even if one assumes that a global SoS estimator may be used to predict such constant SoS value, it is seen that this ``best'' SoS value is not consistent across different experiments, thus no constant value would be ideal. 

It is seen that in all the experimental configurations the mean absolute difference $D$ is the least when MA-PWI data is beamformed using a local SoS map $\sigma_{map}$, with the reported differences being below the reported standard deviations, i.e.\ the experimental noise floor.
It is also worth to note that the SWS values estimated using $\sigma_{map}$ are relatively consistent across the experimental setups as well, although these vary largely for the values estimated using constant SoS beamforming.
In addition, from the reported standard deviations, it is observed that SWS values exhibit more variations closer to the ARF push, with similar observations reported earlier by us in~\cite{otesteanu2019spectral,chintada2020nonlinear} as well as by other groups in~\cite{bernard2016frequency,kijanka2019two}.

In this work, we corrected the phase aberrations in the received MA-PWI data but not in transit delays for correcting phase aberrations of ARF pushes.
Therefore the observed biases in SWS estimation are due to the cumulative effect of both phase aberrations in generation and tracking shear-waves. 
Nevertheless, in Table\,\ref{results_tab} the SWS values reported in the CIRS phantom beneath the water and muscle sides are seen to be consistent between experiments \mynum{1} and \mynum{2}, despite the ARF pushes having used fixed Tx delays while being applied through mediums with different SoS.
Similar observations can be made from experiments \mynum{3} and \mynum{4}, indicating that phase aberrations in ARF push generation potentially affect the generated and observed SWS minimally.
These observations also corroborate the findings in~\cite{huang2016phase} as well as our preliminary experiments where incorrect SoS values in ARF generation did not result in significant differences in SWS estimation.
It is however unclear if such aberrations may cause differences in other SWE applications, such as frequency-dependent, nonlinearity, or attenuation measurements, e.g., by changing the spectral content of induced shear-waves.
Nevertheless, one could also correct ARF Tx delays using local SoS maps $\sigma_{map}$, if this becomes relevant for a given application scenario.
Aberration-based SWS errors observed for tracking with PWI are expected to also be present if focused beams are used, since PW and focused beams travel through similar tissue regions, especially in a layered medium.
These SWS errors should similarly be correctable using local SoS adaptive beamforming.
Note that in our beamforming of MA-PWI, we are already utilizing Rx-focusing, where the adaptive beamforming is shown to be advantageous, so given reciprocity with time-reversal, one would expect such benefit to also exist for focusing on the Tx side.

\begin{figure}
    \centering
    \includegraphics[width=\textwidth]{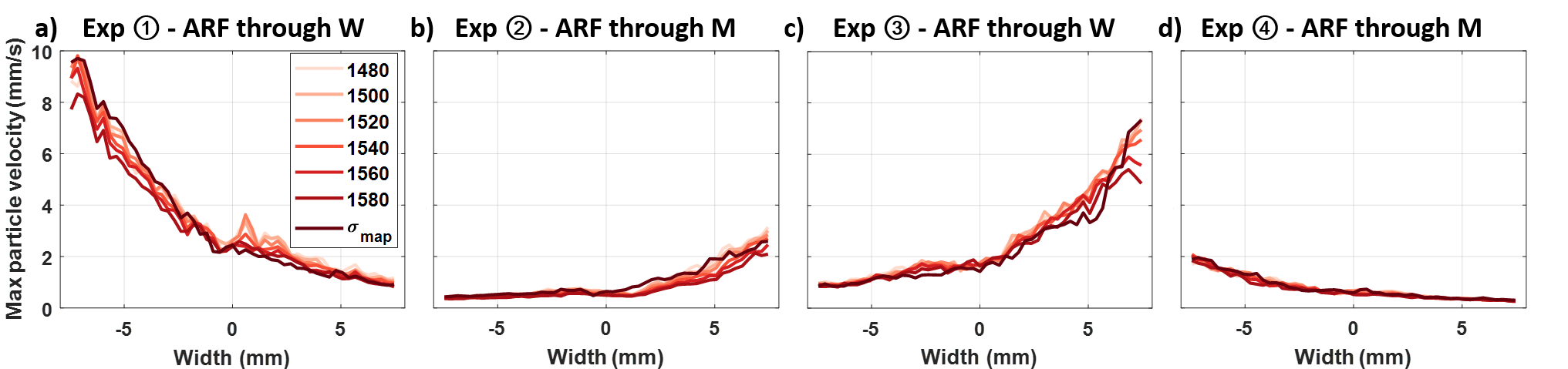}
    \caption{Maximum shear-wave particle axial velocity values for the four experimental configurations, with each line showing the velocity observation from the same MA-PWI data but using different SoS assumptions in beamforming, i.e. constant values from 1480\,m/s to 1580\,m/s vs.\ a spatial map $\sigma_{map}$.}
    \label{fig:max_axial_velocity}
\end{figure}

To qualitatively illustrate the effect of aberration on shear wave amplitudes, when ARF is focused through different materials and MA-PWI is beamformed with different SoS values, the mean maximum amplitude of axial velocity profiles within the same ROI that was used for plotting axial velocity profiles in Figure\,\ref{fig:results}(\RomanNumeralCaps{1}) are shown in Figure\,\ref{fig:max_axial_velocity} for all the experimental configurations.
The average of maximum shear-wave particle velocities in experiments \mynum{2} and \mynum{4} with ARF pushes through the muscle are seen to be $\approx$3.8 times smaller compared to that of experiments \mynum{1} and \mynum{4} with pushes through water, given that ARF focusing was performed with a fixed assumed SoS. These findings corroborate the observations in \cite{carrascal2016phase}. In contrast, differences in shear-wave particle velocities due to beamforming with different SoS values is seen to be minimal for all experimental configurations.

We herein show that local SoS based adaptive beamforming corrects phase aberration effects on SWS estimation. To that end, robust and accurate local SoS estimation is key. Several groups showed SoS reconstruction for submersible body parts using ring or rotating transducer setups.
To alleviate the need for complex setups, hand-held solutions with acoustic reflector based, e.g.~\cite{sanabria2018speed,chintada2021time}, and pulse-echo disparity based, e.g.~\cite{sanabria2018spatial,rau2021speed}, tomographic reconstruction methods were also demonstrated.
To achieve fast, real-time SoS map estimations, such reconstructions have also been accelerated using deep-learning based techniques in~\cite{vishnevskiy2019deep} and \cite{bernhardt2020training}.

\subsection{Simulation study}
To better reason about the observations above, we conduct a simulation experiment.
For a simulation of deformations caused by shear-waves and the imaging thereof, one would require a complex continuum mechanics simulation that can faithfully model the ARF, tissue dynamics, and the imaging interactions, each of which requiring very complex and mostly unknown parameterizations.
Not only setting up such a simulation would be very difficult, but also any results would be highly sensitive to chosen parametrizations and thus potentially questionable.
Therefore, we chose herein a simplified simulation scheme, where we neglect the ARF mechanics and energy exchange, as well as the shear-wave dynamics, and focus instead on the effects from SoS in beamforming.

This is also in line with our observations in Figure\,\ref{fig:max_axial_velocity} that although the maximum particle velocities and hence the displacement amplitudes may vary, mainly due to ARF push mechanics, the estimated SWS nevertheless is relatively independent of such amplitudes and thus ARF mechanics, and is largely dependent on the beamforming (SoS) as seen from the rest of our results.

Accordingly, we focus on an already-generated shear-wave and simulate the SWS estimation process instead.
Most forms of SWS estimation operates on tracked displacement-time profiles (DTPs) as an observation of the passing shear wavefront, as illustrated in Figure\,\ref{fig:results}(\RomanNumeralCaps{2}).
Methods may use cross-correlation of such DTPs at known spatial increments (variable $p$ in our method), or even simple peak-picking in DTPs (e.g., the black dots in Figure\,\ref{fig:results}(\RomanNumeralCaps{2},d)), in order to infer the time elapsed for the wave to travel this distance.
In our simulations, we place scatterers (``anchors'') to mark the physical locations of regular intervals where the shear-wave generates such DTPs.
We then look at the beamformed images to see where the anchors (scatterers) are observed, which will also be the location where the DTPs will be recorded from an SWS-estimating observer point-of-view.

\begin{figure}
\centering
\includegraphics[width=1\textwidth]{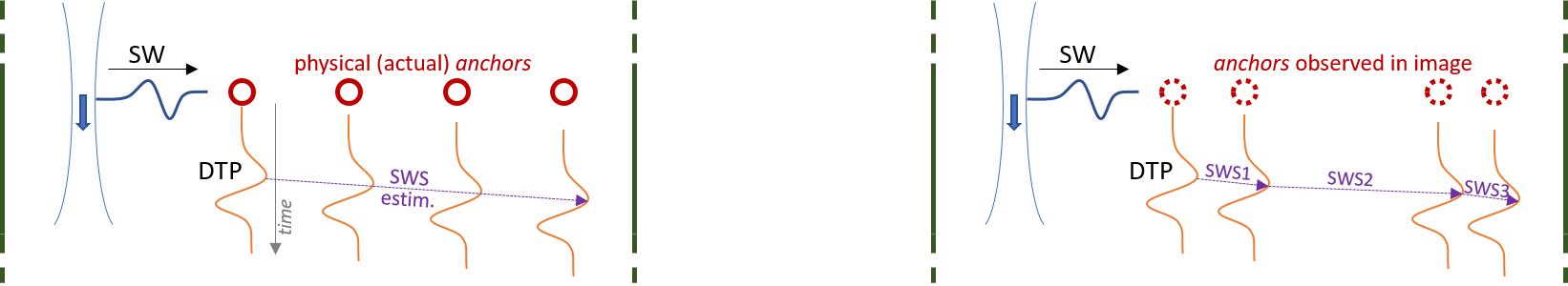}
\caption{Illustration of simulation. A given SWS encountered as displacement-time profiles (DTPs) at physical anchor (scatterer) location (left) may appear as a different SWS if the same DTPs are observed at different image locations (right), e.g., due to aberrations in beamforming.}
\label{fig:sim_illustration}
\end{figure}

We use the SoS map $\sigma_{map}$ in Figure\,\ref{fig:exp_configurations}(c) in the k-Wave acoustics simulation toolbox~\cite{treeby2010k}.
To mark physical tissue anchor points along a laterally propagating wavefront with a constant speed, we place a simulated point scatterer at a fixed depth of $z=27.3$\,mm, while changing its lateral location
from $x=-13.8$\,mm to $+13.8$\,mm, at intervals of 0.3\,mm, leading to 93 individual scatterer locations.
For each scatterer location, an MA-PWI is simulated with three angles  \{-8, 0, 8\}$^{\circ}$ was acquired, each beamformed using delay-and-sum assuming a homogeneous medium with constant SoS of 1580\,m/s, and these triplet images were then spatially compounded as in our standard SWS imaging approach above.
In each of the 93 beamformed and compounded frame, we then easily identify the peak of the isolated scatterer (based on envelope intensity and a quadratic subsample approximation~\cite{azar2010sub}) and treat this apparent anchor location as the \emph{observed} location of a DTP which actually occurs in the physical location of the anchor, as seen in Figure~\ref{fig:sim_illustration}.
Assuming an MA-PWI framerate of 10K frames/s, 0.3\,mm physical anchor distance per tracking frame time of 100\,$\mu$s corresponds to a 3\,m/s shear-wave travel.

\begin{figure}
\begin{center}
\includegraphics[width=1\textwidth]{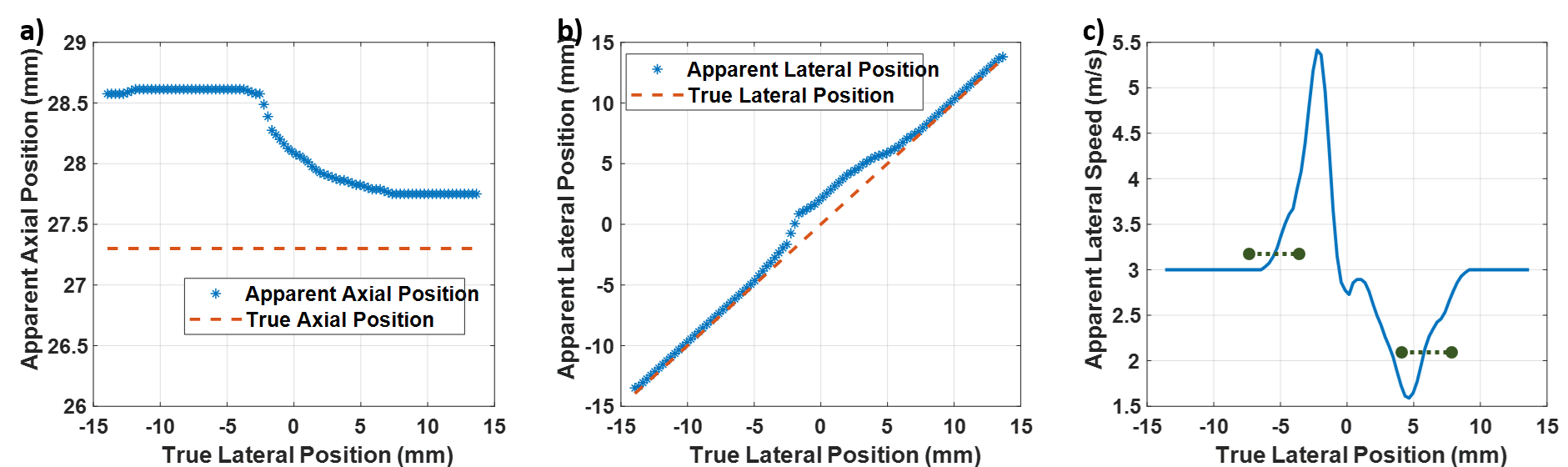}
\end{center}
\caption{Simulation results. True vs.\ apparent axial (a) and lateral (b) positions of a scatterer, observed via beamforming with a homogeneous SoS map of 1580\,m/s. Apparent lateral speed~(c) computed using the estimators similarly to the experimental case. The black markings represent the mean estimated SWS within the width of the two ROIs used in the experiments and delineated in Figure\,\ref{fig:results}(\RomanNumeralCaps{3},b)}
\label{fig:k_wave_simulation}
\end{figure}

In Figure\,\ref{fig:k_wave_simulation}(a),
we plot the true axial position given in the simulation against the apparent axial-axis position measured from the beamformed data.
As can be seen, underneath the muscle tissue (the SoS of which is closer to the chosen constant beamforming SoS) the apparent depth of the scatterer is closer to the actual depth.
Note that a near-vertical planar shear wavefront, such axial shift would not have a major effect.
To gain insight into lateral propagation characteristics, we also plot the true lateral positions against the apparent lateral apparent position in Figure\,\ref{fig:k_wave_simulation}(b).
One can see there the offset from the true lateral positions in a large transition region where some but not all beamforming paths cross the aberrator.
This then perturbs the SWS observed from these observations.
An extreme example, for instance, the lateral observations of two consecutive locations $x=\{-2.25,-1.95\}$\,mm that are then observed as the apparent location $x=\{-1.35,0.75\}$\,mm, which using some finite-difference speed estimation hence would distort the actual shear-wave speed of 3\,m/s, to locally appears as 21\,m/s in the beamformed data.
For SWS estimation in practice, instead of finite-differences, often some finite filter length is used for noise suppression.
In Figure\,\ref{fig:k_wave_simulation}(c), we converted the lateral position observations to SWS estimation using a smoothing approach similar to the practical phantom experiment earlier.
This figure illustrates how SWS is inflated or deflated due to SoS inhomogeneity effects although the actual physical SWS is 3\,m/s. 
In this figure we mark the regions corresponding to width of the left and right ROIs reported earlier in our experiments, with the marked lines indicating the mean SWS estimation in these regions.
From this simulation result, a slight overestimation of SWS under the water side as well as a gross underestimation under the muscle side are observed.
These results fully corroborate the findings in the experiments in Figure\,\ref{fig:results}(\RomanNumeralCaps{4}).

Note that if we see the anchors at different locations in the beamformed image, we would also observe a corresponding DTP at that different location in a SWS tracking scenario.
Therefore, a cross-correlation yielding the same time delay would be observed at a different apparent distance, hence resulting in a different observed SWS.
For instance, if two neighbouring anchors are to be observed laterally 0.5\,mm apart, since the cross-correlation of their DTPs would yield a time lapse of 100\,$\mu$s, we would observe an SWS of 5\,m/s, which describes the value observed around x=-2.5\,mm in the simulation results in Figure\,\ref{fig:k_wave_simulation}.

By representing the observed DTP locations with a scatterer, we omit the effects of SoS in the displacement estimation itself.
This is based on the assumption that the SoS-related beamforming delays can differ only negligibly across the spatial range of shear-wave related particle displacements.
Furthermore, any delay differences that may skew the amplitude of observed particle velocities are likely to affect the SWS estimation negligibly, as also observed from the particle amplitudes reported in Figure\,\ref{fig:max_axial_velocity} being unrelated to our resulting SWS estimations.
In our simulations, we ignored DTP distortions due to aberrations, i.e.\ cross-correlation leading to any artifactual time shifts in addition to spatial shifts.
Such distortions occur, e.g.\ beneath refracting edges as seen within the $\pm$1\,mm band in the DTPs in Figure\,\ref{fig:results}(\RomanNumeralCaps{2}).
Nevertheless, such distortions often lead to random noise with non-systematic offset and to very low correlation coefficients with SWS estimation being inconclusive in these distorted regions.
These assumptions need to be further studied in the future with a comprehensive full-pipeline simulation study.

\section{conclusions}
In this work, SoS aberration effects on SWEI when beamformed with different SoS have been studied, with ARF pushes and shear-wave observations through water and muscle samples.
We found that ARF push generation affect the generated as well as observed SWS minimally, while the beamforming errors and aberrations may significantly alter the observed SWS values. 
Average SWS disparity when a constant SoS value is used for beamforming was found over 4.3 times worse than using a known SoS map. 
Indeed there was no single constant SoS value that could mitigate the observed SWS biases.
Nevertheless, it is shown that using a known SoS map in beamforming delay calculations prevents such SWS estimation biases.
Note that even using only direct linear paths in beamforming, i.e.\ not taking potential refractions into account, major SWS errors could be prevented.

Without aberration correction, the observed differences in measured SWS were substantial.
In terms of elastic moduli, e.g.\ computed given the SWS values in Table\,1 at 1540\,m/s and assuming density of 1\,g/cm$^3$, our results show that for the same material one may find 7.56 vs.\ 4.67\,kPa for experiment \mynum{1} and 8.94 vs.\ 5.15\,kPa for \mynum{3}.
These are important differences which may cause misdiagnosis, especially given that absolute values of SWS measurements are typically used in staging diseases in the liver, breast, and kidney~\cite{sarvazyan2011elasticity,deffieux2015investigating,tanter2008quantitative}.
These large differences indicate the difficulty in standardizing SWS measurements and that diseases may be staged incorrectly which can impact management and treatment decisions.
For instance, with 7\,kPa being the common threshold between mild (F1) and significant (F2) fibrosis, a patient may be staged differently given the pairs of measurements above.
It needs to be further studied if such measurement errors would scale linearly or stay constant at increasing SWS values.

Since soft tissues are intrinsically inhomogeneous, for accurate tissue characterization and diagnosis using SWEI, it appears imperative given our study to beamform MA-PWI data using accurate SoS distributions of the medium to alleviate possible confounding effects of SoS and beamforming on estimated SWS. 
For estimating such local SoS distributions, one could use the same ultrasound transducer used for SWEI as in~\cite{jaeger2015computed,sanabria2018spatial,rau2019speed} as demonstrated for aberration correction in~\cite{rau2019aberration}. 

\section{Acknowledgements}
Funding was provided by the Swiss National Science Foundation (SNSF) and the Promedica Foundation, Chur.

\bibliographystyle{unsrtnat}
\bibliography{main}
\end{document}